\documentclass[fleqn,usenatbib,useAMS]{mnras}

\usepackage{graphicx}   
\usepackage{amsmath}    
\usepackage{amssymb}    
\usepackage{multicol}        
\usepackage{bm}         
\usepackage{pdflscape}  

\newcommand{\hmsun}{h^{-1}{\rm ~M}_\odot}
\newcommand{\hmpc}{h^{-1}{\rm ~Mpc}}

\newcommand{\kms}{{\rm ~km/s}}

\usepackage[T1]{fontenc}
\usepackage{ae,aecompl}

\usepackage{txfonts}

\title[large--scale velocity flows]{Voids and Superstructures:\\ correlations and induced large--scale velocity flows}

\author[Lares et al.]{\parbox[t]{\textwidth}{%
Marcelo Lares\thanks{Contact e-mail:
   \href{mailto:marcelo.lares@unc.edu.ar}{marcelo.lares@unc.edu.ar}}, 
Heliana E. Luparello, Victoria Maldonado, Andr\'es N. Ruiz, Dante J. Paz,
Laura Ceccarelli, \& Diego Garcia Lambas}\vspace{0.2cm}\\ 
Instituto de Astronom\'{\i}a Te\'{o}rica y Experimental, CONICET-UNC, 
and Observatorio Astron\'{o}mico de C\'{o}rdoba, 
Argentina}

\date{Released 2016 Xxxxx XX}
\pagerange{\pageref{firstpage}--\pageref{lastpage}} \pubyear{2016}

\def\LaTeX{L\kern-.36em\raise.3ex\hbox{a}\kern-.15em
    T\kern-.1667em\lower.7ex\hbox{E}\kern-.125emX}

\begin{document}
\label{firstpage}
\maketitle

\begin{abstract} 
The expanding complex pattern of filaments, walls and
voids build the evolving cosmic web with material flowing from underdense
onto high density regions. 
Here we explore the dynamical behaviour of voids and galaxies in void
shells relative to neighboring overdense superstructures, using the
Millenium Simulation and the main galaxy catalogue in Sloan Digital
Sky Survey data.
We define a correlation measure to estimate the tendency of voids to
be located at a given distance from a superstructure. 
We find voids--in--clouds (S--types) preferentially located
closer to superstructures than voids--in--voids (R--types)
although we obtain that voids within
$\sim40\hmpc$ of superstructures are infalling in a similar fashion
independently of void type.
Galaxies residing in void shells show infall towards the closest
superstructure, along with the void global motion, with a differential
velocity component depending on their relative position in the shell
with respect to the direction to the superstructure.
This effect is produced by void expansion and therefore is stronger
for R--types.  
We also find that galaxies in void shells facing the superstrucure flow
towards the overdensities faster than galaxies elsewere at the same
relative distance to the superstructure.
The results obtained for the simulation are also reproduced for the
SDSS data with a linearized velocity field implementation.
\end{abstract} 


\begin{figure} 
   \centering
   \includegraphics[width=0.47\textwidth]{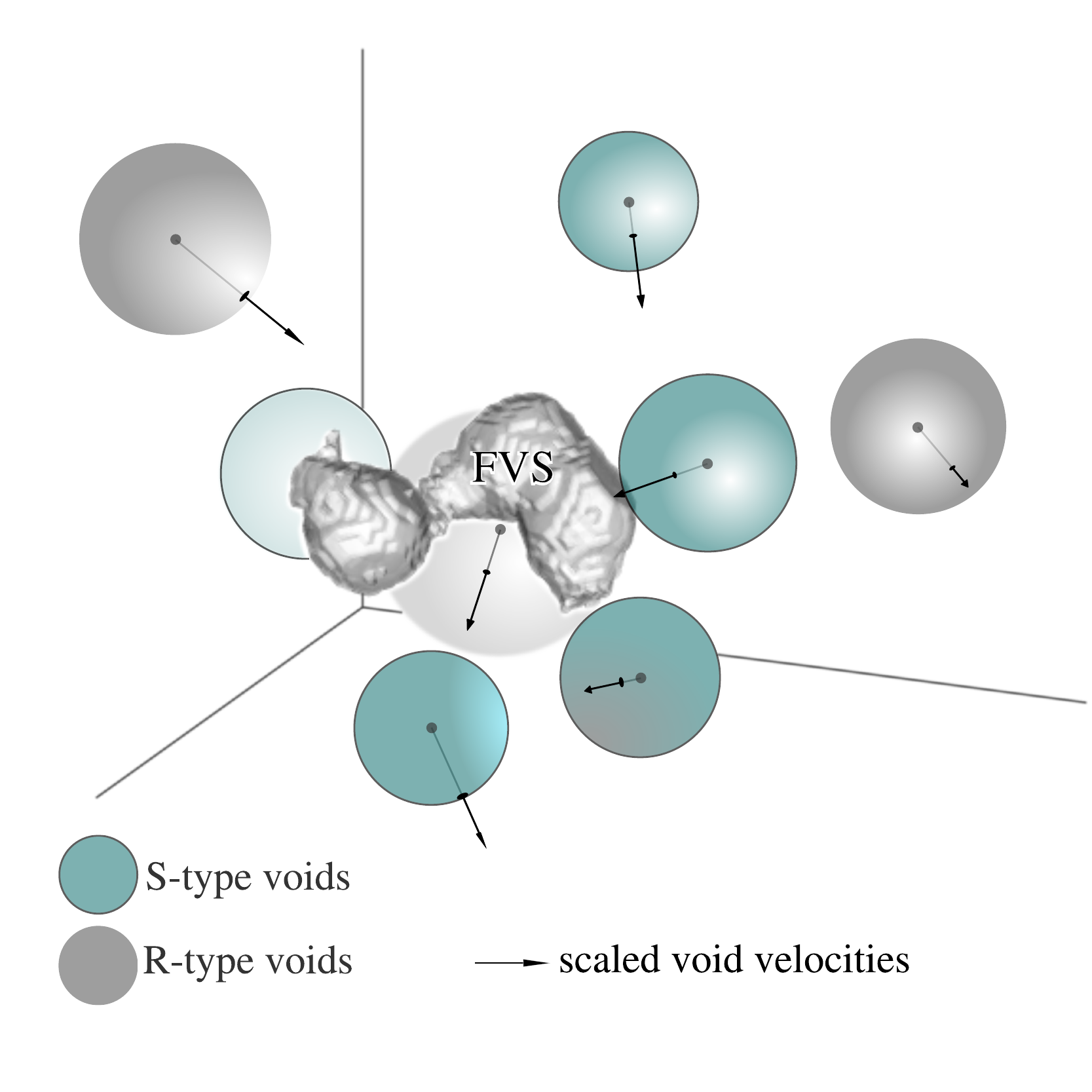} 
   \caption{A Visualization of a group of void--FVS pairs, including
      all the voids (represented as spheres) that have the FVS (in the
      center of the Figure) as their closest one.  Arrows are scaled
      representations of the void velocity vector, with dotted
      segments located inside or behind voids.  This example
      corresponds to a subset of data from the simulation.   Other
      FVSs and voids in the neighbourhood are not shown for
   simplicity.  Darker spheres represent R--type voids and delineated
spheres represent S--type voids.   The latest are typically closer and
moving towards the FVS.} 
   \label{fig:figure1} 
\end{figure}

\begin{figure} 
   \centering
   \includegraphics[width=\columnwidth]{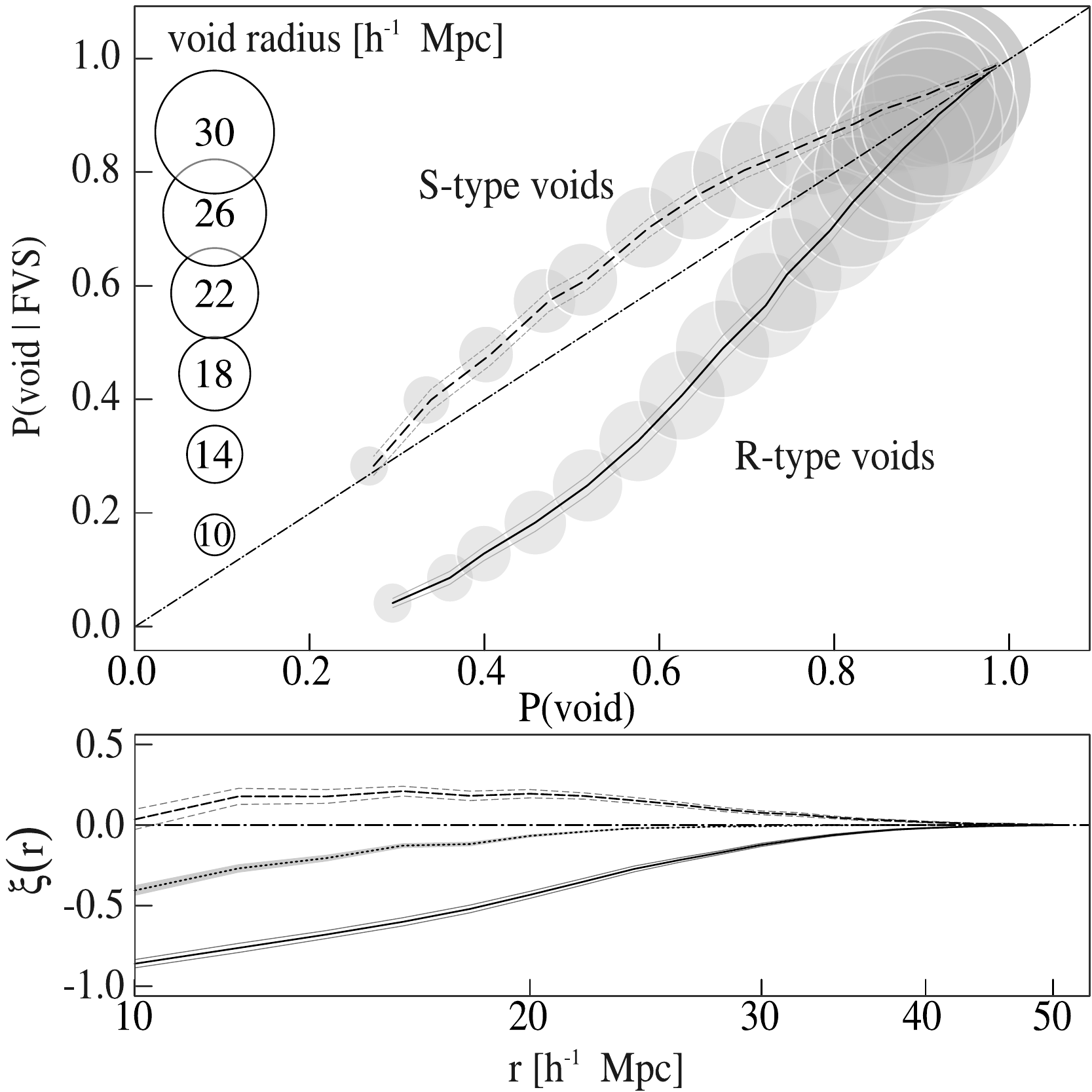} 
   \caption{Spatial distribution of voids with respect to FVSs, 
      in the simulation.    
      \textit{Upper panel:} 
      The curves show the conditional probability, $P(void | FVS)$, of
      finding a randomly placed sphere containing simultaneously void
      and FVS volume fractions, as a function of the probability
      $P(void)$ that the sphere contains a fraction of void volume, in
      spite of the presence of FVS.  If the locations of voids and
      FVSs were independent, $P(void | FVS) \simeq P(void)$ for any
      value of sphere radius (dot--dashed line indicating a
      one--to--one relation).  The upper curve corresponds to the
      results for S--type voids, and the lower curve to R--type voids.
      Circle radii scale linearly with the radius of the random spheres.  
      Uncertainties are computed using Jackknife resamplings.
      \textit{Bottom panel:} Void--FVS correlation function $\xi(r)$.
      We use the definition of $\xi(r)-\xi_{\rm void-FVS}(r)$ given in Eq.
      \ref{E_P4}.
      S--type voids exhibit a positive correlation in a wide range of 
      scales, consistent with their denser environment.    On the other 
      hand, R--type voids and FVSs are anticorrelated, with a stronger 
      signal than that of S--type voids.
      The central curve corresponds to the full sample of voids,
      without separating by type.
   }
   \label{fig:figure2}
\end{figure}

\section{MOTIVATION} \label{S_intro}

The cosmic web is the largest scale outcome of the anisotropic growth
of mass overdensities.
It also represents the transition between the linear and non-linear
regimes and encodes information about the early phases of structure
formation.
The analysis of the mass transport between different environments
clearly shows how matter flows from voids into walls, and then via
filaments into cluster regions, which form the nodes of the cosmic web
\citep[see e.g.][]{vandeweygaert_cosmic_2011, cautun_evolution_2014},
producing different redshift evolution of haloes in different
environments \citep{hann_evolution_2007}.

The relations between the different types of the largest structures in
the universe have been suggested on several contexts
\citep{einasto_structure_1986, einasto_supercluster_1997,
   platen_alignment_2008, aragoncalvo_spine_2010,
einasto_towards_2011}.
This allows to analyse the large--scale structure in terms of the
large overdense structures, or alternatively in terms of the large
underdense regions.
Actually, the largest overdense structures that shape the cosmic web,
also serve as boundaries for voids
\citep{cautun_evolution_2014}.
Also, by understanding the evolutionary processes of the different
structures we can obtain a deeper insight on the origin and history of
the present structure
\citep{leclercq_bayesian_2015}.

In order to deepen our understanding on the nature of voids and the
evolution of their properties it is crucial to take into account the
surrounding environment where they are embedded
\citep{paranjape_more_2012}. 
Essentially, the hierarchy of voids arises by the assembly of mass in
the growing nearby structures.
\citet{sheth_hierarchy_2004} suggest that while some voids remain as
underdense regions, other voids fall in on themselves due to the
collapse of dense structures surrounding them.
According to this scenario, void evolution exhibits two opposite
processes, expansion and collapse, being the dominant process
determined by the global density around the voids.
The distinction between these two types of void behaviour depends on
the surrounding environment.
It is expected that the large underdense regions with surrounding
overdense shells will undergo a {\it void-in-cloud} evolution mode.
These voids are likely to be squeezed as the surrounding structures
tend to collapse onto them.
On the other hand, voids in an environment more similar to the global
background density will expand and remain as underdense regions
following a so called {\it void-in-void} mode.
In a series of works
\citep{ceccarelli_clues_2013, paz_clues_2013, ruiz_clues_2015} we have
considered an alternative classification of void according to their
environment taking into account the cumulative radial density
profiles.
Voids with a surrounding overdense shell exhibit a rising
cumulative density profile, which overcompensates the underdensity and
therefore reach possitive values between 2 and 3 times the void
radius.
These voids, dubbed S--type after its shell--like structure, are
likely collapsing.
On the other hand, voids with a smoothly rising integrated
density profile that approaches the mean density at large distances
(hereafter R--type voids) show a continuos expansion.
In this scheme, R-type voids resemble void-in-void objects while void
in clouds are consistent with the S-type definition.

The origin of the large--scale flows observed in the local Universe
has been subject of controversy during the last decades.
While several works have focused on the search of a great attractor
consistent with peculiar velocities of local structures, an
alternative approach includes the expansion of the local void as
proposed by \citet{tully_localvoid_2008, tully_laniakea_2014}. 
Albeit the contribution of the local void dynamics improves the
description of the velocity fields in the nearby Universe, the
estimated peculiar velocity of the local group is still conflicting
with that predicted from the infall onto the Shapley supecluster and
the local void expansion, due to the presence of residual velocities
of $\sim 200\kms$ for the local group \citep[][and references
therein]{vandeweygaert_voids_2016}.

In recent works we have reported the non-negligible motions of voids
as a whole \citep{ceccarelli_sparkling_2016, lambas_sparkling_2016},
adding a new component affecting galaxy peculiar velocities.
These global motions can be a key piece to complete the scenario of
the dynamics of local structures. 
In this work we show that the velocities generated by the mass
under/over densities associated with the large--scale structures of
the galaxy distribution are complementary to obtain a detailed
description of the large--scale velocity flows.
In this context, we study the joint dynamics of galaxies with respect
to voids and Future Virialized Structures (hereafter FVSs), in order
to explore coherent patterns of motions of mass from the shells of
voids to the largest overdensities.

In the next section, we describe the general methods used to identify
voids and FVSs, both in semianalytic galaxies and in an observational
galaxy catalogue.
Then, in Sec. \ref{S_results} we analyse the spatial distribution of
voids relative to superstructures and their associated dynamics.
The same analysis is then performed to galaxies, where their motions
are considered separately for different configurations of relative
positions of galaxies, voids and FVSs.
In Sec. \ref{S_observations} we also present similar studies applied
to observational data.
Finally, we present a discussion of our results in the context of
recent works in Sec. \ref{S_conclusions}.


\section{DATA} \label{S_methods}

\subsection{Galaxy catalogues}
\label{SS_galaxy_catalogues}

The observational galaxies used in this work correspond to the Main
Galaxy Sample of the Sloan Digital Sky Survey Data Release 7
\citep[SDSS-DR7,][]{abazajian_dr7_2009}. 
This catalogue comprises almost a million of spectroscopic galaxies
with redshift measurements up to $z \le 0.3$ and an upper apparent
magnitude limit of 17.77 in the $r$-band.
From this sample we select galaxies with a limiting redshift $z=0.12$
and a maximum absolute magnitude in the $r$-band of $M_r-5\log_{10}(h)
= -19.95$, in order to obtain a volume complete sample of galaxies at
that redshift.
The limits of this sample are chosen on the basis of a compromise between the volume and the number of tracers.
For SDSS velocities, we adopt the peculiar velocity field presented by
\citet{wang_reconstructing_2012}. 
The authors used the linear theory connection between mass overdensity
and peculiar velocity to reconstruct the 3D peculiar velocity field of
SDSS galaxies \citep{wang_reconstructing_2009}.
With this velocity field we compute the bulk void velocities in the
SDSS sample.
For an analysis of the effects of using linearised velocities, we
refer the reader to the Appendix of \citet{ceccarelli_sparkling_2016}.
We also analysed the influence of large--scale flows in observational
data.
For this purpose we used a galaxy group catalogue which follows the
identification method presented in \citet{yang_halo_2005}, applied to
SDSS-DR4 in \citet{yang_galaxy_2007}, and updated by the authors to
the SDSS-DR7\footnote{http://gax.shao.ac.cn/data/Group.html}.
For this sample of groups, we also adopted the linearized velocity
field by \citet{wang_reconstructing_2012}.

On the other hand, the simulated galaxies used were extracted from a
semi-analytical model of galaxy formation (SAM) applied to the
Millennium Simulation \citep[MS,][]{springel_ms_2005}, a cosmological
$N$-body simulation which counts with $2140^3$ dark matter particles
evolved in a cubic comoving box of $500\hmpc$ on a side. 
The cosmological parameters adopted for the MS correspond to
\textit{WMAP}1 results \citep{spergel_wmap1_2003}, i.e. a flat
$\Lambda$ cold dark matter cosmology with $\Omega_{\rm m}=0.25$,
$\Omega_\Lambda = 0.75$, $\Omega_{\rm b} = 0.045$, $\sigma_{\rm 8} =
0.9$, $h=0.73$ and $n=1.0$.
Given that the MS is a dark matter only simulation, their dark matter
haloes needs to be populated with a SAM to obtain a galaxy population.
In this work we use the public catalogue developed by
\citet{guo_sam_2011}, which is available at the Millennium
Database\footnote{http://gavo.mpa-garching.mpg.de/Millennium}.
In order to make a fair comparison between observations and simulated
data for the complete simulated galaxy population, we select a sample
with the same number density than the observations
This was achieved by selecting all galaxies brighter than \mbox{$M_r -
5\log{h} = -20.6$}, which guarantees the volume density required. 
Althought this threshold in absolute magnitude is not the same than
the limiting magnitude used to define the volume--limited sample of
galaxies, it reproduces satisfactorily the distribution of galaxies.
\citep[see e.g.][]{contreras_density_2013,contreras_density_2015}.
The difference between the two values arises from the fact that
semianalytical models does not reproduce exactly the luminosity
function of galaxies in SDSS.
 

\subsection{Catalogues of cosmic voids}
\label{SS_voids_catalogue}

To construct the void catalogue we follow the procedure described in
\cite{ruiz_clues_2015}, which is a modified version of the algorithms
presented in \cite{padilla_spatial_2005} and
\cite{ceccarelli_voids_2006}.
The galaxies (either simulated or observed) are the tracers of the
density field used in void the identification. 
The method starts with a contrast density field estimation performed
with a Voronoi tessellation, where underdense cells are selected as
void candidates.
Centred in those positions, we compute the integrated density contrast
$\Delta(r)$ at increasing values of radius $r$, selecting as void
candidate the largest spheres which satisfy the condition
$\Delta(R_{\rm void})<-0.9$ and defining $R_{\rm void}$ as the void
radius.
Afterwards, the centre position of void candidates are randomly tilted
and the sphere is allowed to grow in order to recentre the void.
Finally, a void of radius $R_{\rm void}$ is selected as the largest
sphere satisfying the underdense condition and not overlapping with
any other underdense sphere.
The final catalogues comprises 1676 voids for the full Millennium box
and 495 voids for the total SDSS sample, but for dynamical analyses we
used only 288 of them (121 R--type and 167 S--type), which belong to
the inner region of the survey where the velocity field of
\citet{wang_reconstructing_2012} has a reliable reconstruction.
The void radii are in the range 8-26 $\hmpc$, both in SDSS and the
simulation.

\subsection{Catalogues of superstructures}
\label{SS_FVS_catalogue}

In previous works, we have defined the procedures to select
superstructures that will be virialized systems in the distant future
\citep{luparello_future_2011}.
The criteria applied can be used to both observational and simulated
galaxy catalogues.
The method relies on the computation of the luminosity density field,
by convolving the spatial distribution of galaxies (either observed or
SAM) with a kernel function weighted by galaxy luminosity. 
We adopted an Epanechnikov kernel of 8$\hmpc$ to sample the density
field into a grid composed by cubes of 1$\hmpc$ side. We then select
the highest luminosity density groups of cells to isolate the large
structures that will become virialized systems in the future
\citep{dunner_limits_2006}
In our final catalogue we identified 150 FVSs comprising 11394
galaxies in the SDSS, out of which 105 FVSs are within the region
where we used the reconstructed velocity field data by
\citet{wang_reconstructing_2012}.
The simulation, on the other hand, has 790 FVSs due to its larger
volume, which is approximately 6 times the SDSS volume.



\begin{figure} 
   \centering
   \includegraphics[width=0.47\textwidth]{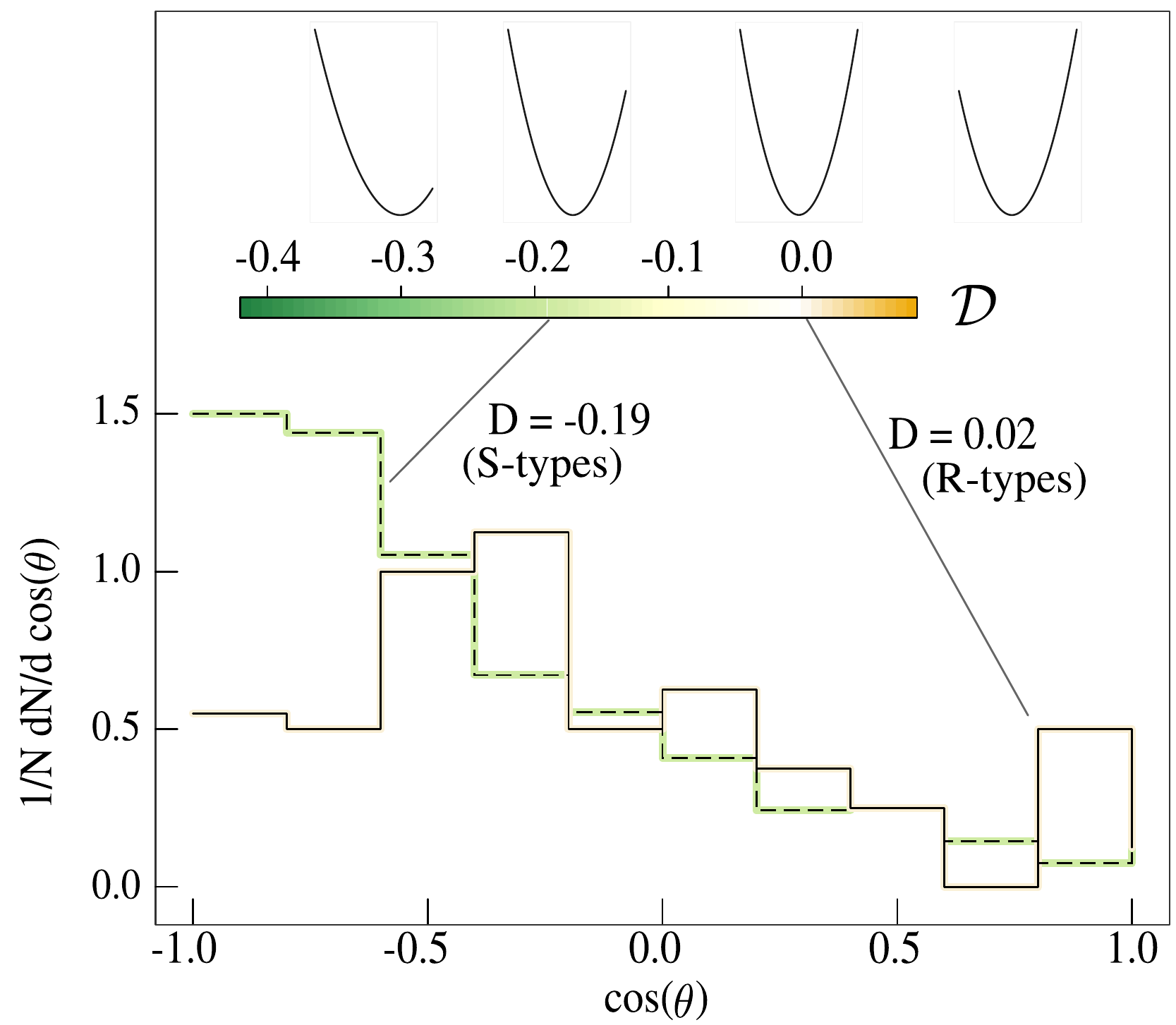} 
   \caption{Normalized histograms of the distribution of
   cos$(\theta)$, where $\theta$ is the angle between the relative
   velocity and the relative position of the void--FVS system (see
   text for details); for subsamples of R--type voids (dotted line) 
   and for S--type voids (solid line) in the simulation.
   These subsamples were chosen in order to show the different
   values of $\mathcal{D}$.
   One sample comprises R--type voids  
   with a separation from the FVS greater than d$>$40$\hmpc$, and
   M$<10^{14} \hmsun$.
   The sample of S--type voids is restricted to d$<$40$\hmpc$, and
   M$>10^{14.2} \hmsun$.                        
   On the top of the figure, we show a scale indicating the
   values of the dipole coefficient $\mathcal{D}$ for both samples.
   We also show, for reference, model distributions for different
   values of the dipole coefficient (-0.4, -0.2, 0, and 0.2,
   respectively).}
   \label{fig:figure3} 
\end{figure}

\begin{figure*} 
  \centering
  \includegraphics[width=0.8\textwidth]{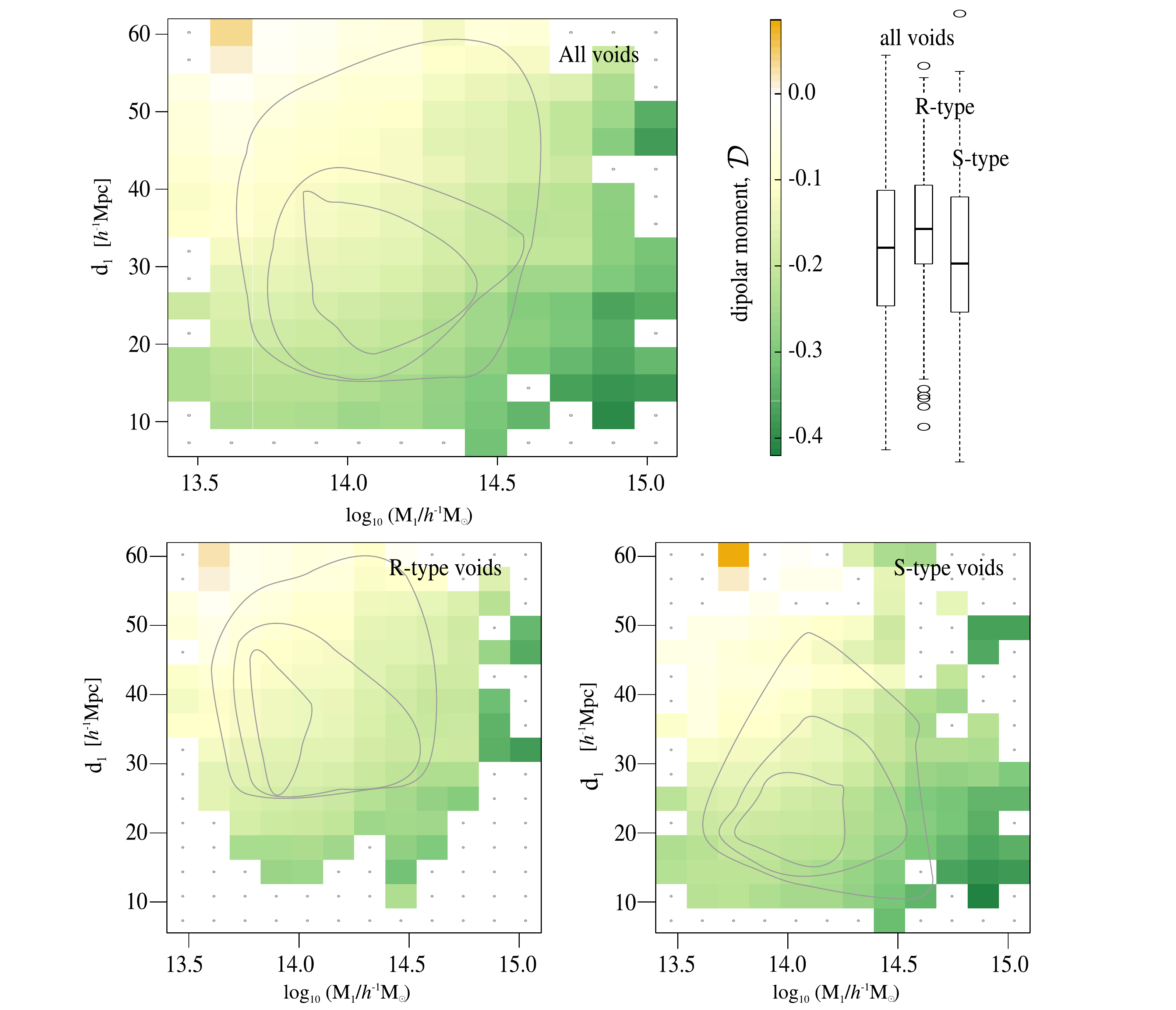}
  \caption{Dependence of the dipole moment on FVS mass and void--FVS
  separation in the simulation data.  
  We show 2D histograms of the dipole moment estimator $\mathcal{D}$
  (encoded in color) for the distribution of cos($\theta$) in bins of
  FVS mass and void--FVS distance.
  Positive values of cos($\theta$) correspond to positive velocity
  components on the FVS--void direction (see text for its definition),
  which indicates that the void is moving away their closest FVS.
  We show in the upper panel the full sample of void--FVS pairs, and
  in the bottom left and right panels, R-- and S--type voids,
  respectively.
  Box plots correspond to the distributions of $\mathcal{D}$ for each
  sample.
  Contour levels are shown for the number of objects contributing to
  the signal in each bin, and correspond to 25, 50 and 75 per cent of
  the pairs.}
  \label{fig:figure4} 
\end{figure*}
             
\begin{figure*} 
  \centering
  \includegraphics[width=0.8\textwidth]{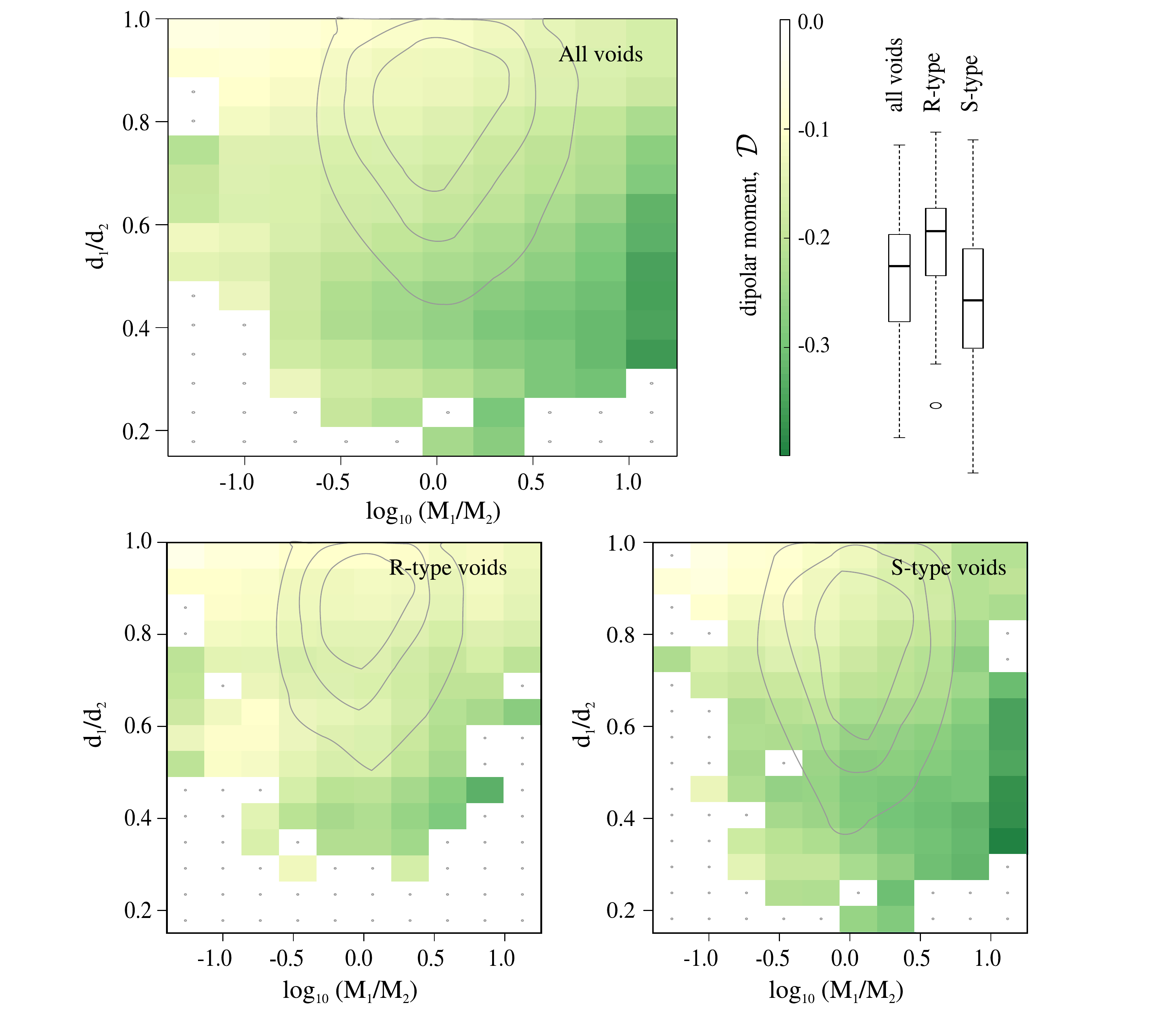} 
  \caption{2D histograms of the dipole moment estimator $\mathcal{D}$
  for the distribution of cos($\theta$), in bins of the ratios of
  masses and distances between the closest and the second closest
  FVS to each void.  This estimator is computed for the closest
  void--FVS pair.  Box plots and contour levels are analogous to those
  of Fig. \ref{fig:figure4}.}
  \label{fig:figure5}
\end{figure*}

\begin{figure}
  \centering
  \hspace{-9pt}
  \includegraphics[width=0.49\textwidth]{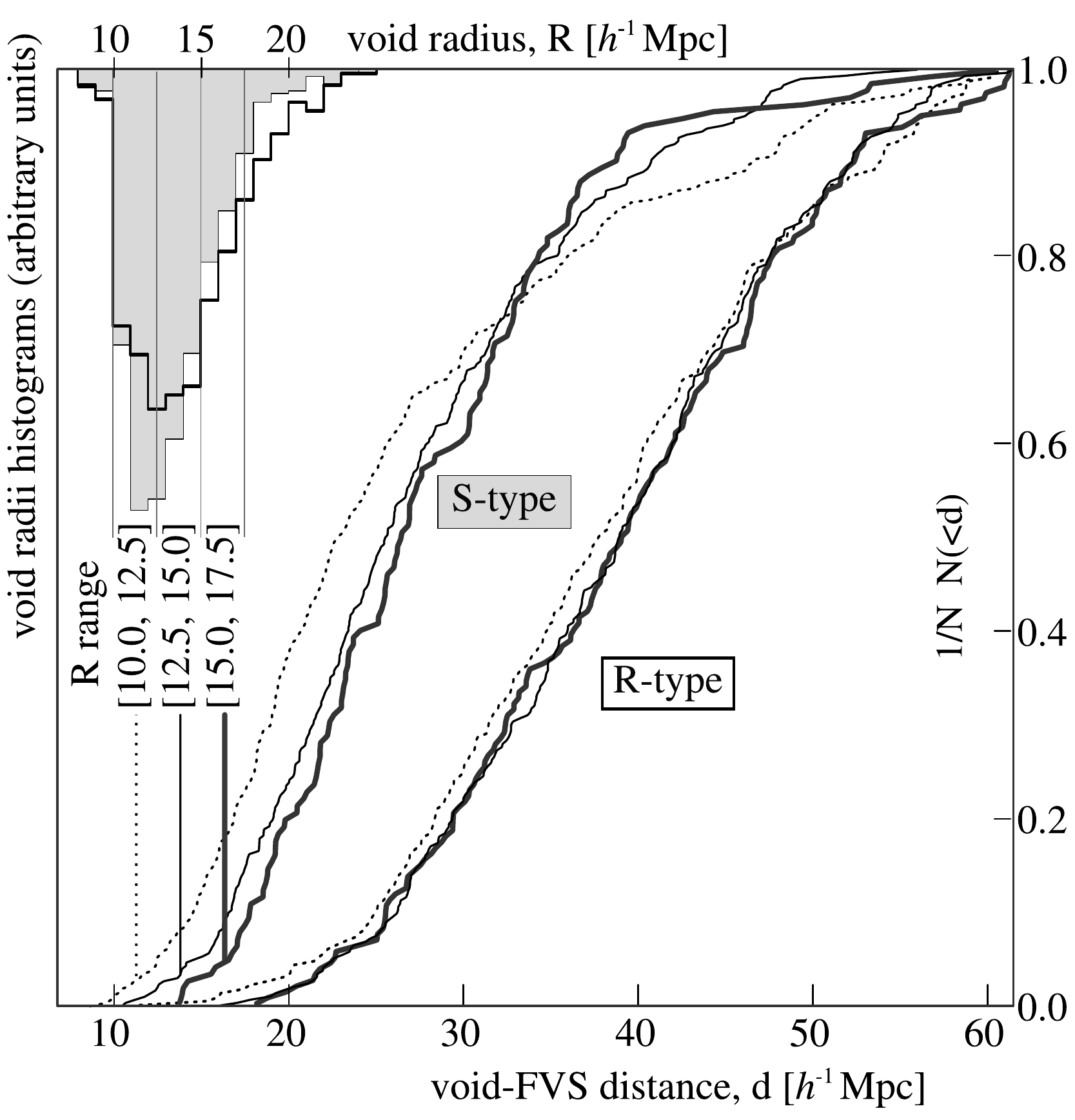}%
  \caption{Cumulative distributions of void--FVS distances (bottom and
     right scales) for R/S--types and for different void radius
     intervals (\mbox{$10.0<R_v[h^{-1}{\rm Mpc}]<12.5$} with dotted
     lines, \mbox{$12.5<R_v[h^{-1}{\rm Mpc}]<15.0$} with thin solid
     lines, and \mbox{$15.0<R_v[h^{-1}{\rm Mpc}]<17.5$} with thick solid
     lines).
     The histograms show the normalized distributions of void radii for 
     R--types (empty bars), and S--types (filled bars).
     It can be clearly seen that
     the difference in the FVS--void distance
     cumulative distributions is dominated by
     void type and not by void size.
  }
  \label{fig:figure6}
\end{figure}
 
\begin{figure} 
   \centering
   \includegraphics[width=\columnwidth]{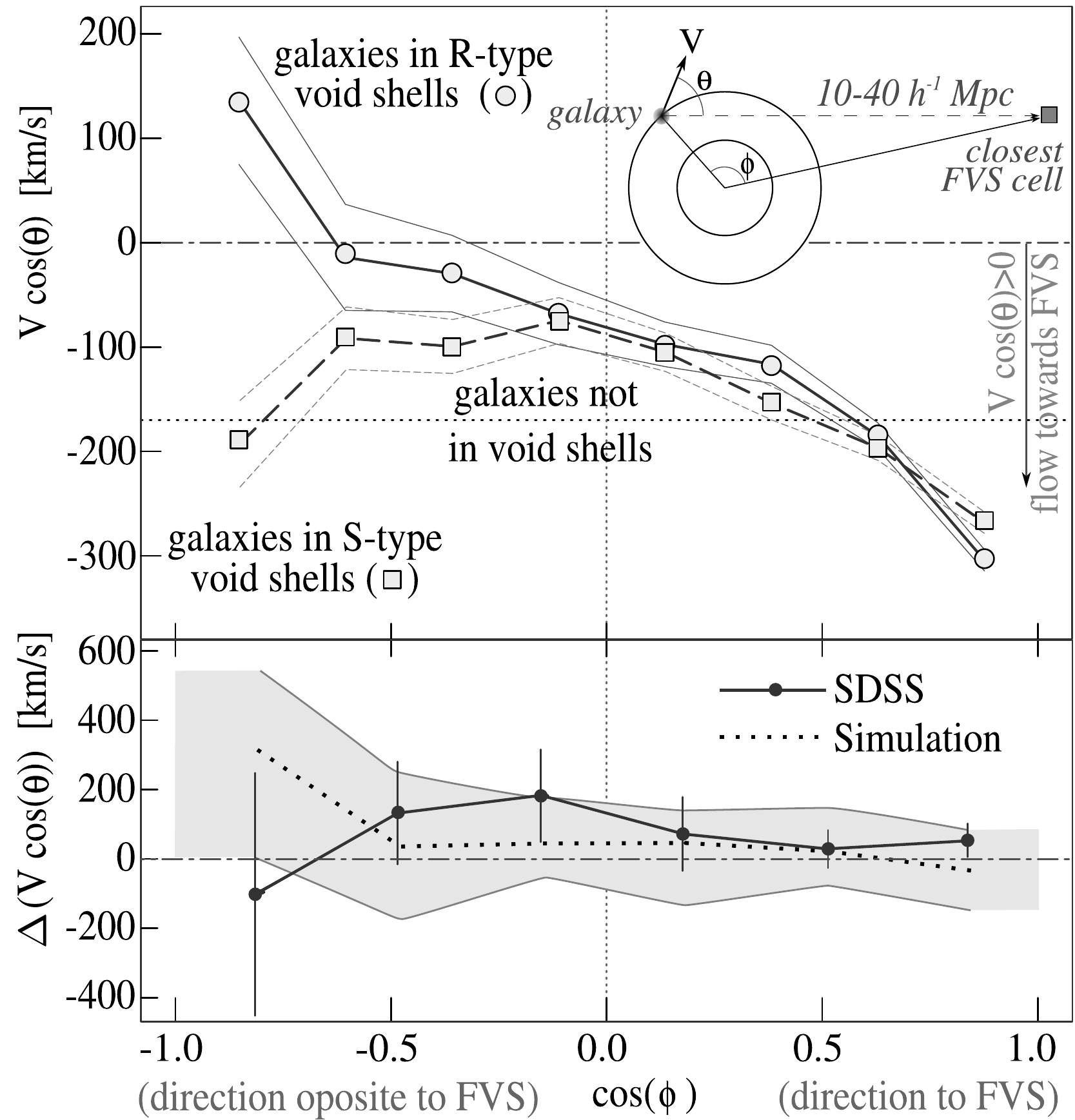} 
   \caption{\textit{Upper panel:} Mean projected velocity, V~cos($\theta$),
   for SAM central galaxies in void shells
   (solid lines:R--types, dashed lines: S--types) along the direction to
   the closest FVS
   as a function of the
   cosine of the angle $\phi$ between the position of the galaxy and the void
   centre--FVS direction.
   SAM galaxies outside voids are shown with an horizontal dotted line.
   \textit{Bottom panel:} Velocity difference 
   between galaxies in void shells of R/S types. 
   The curves show the R$-$S velocity difference of galaxies in void
   shells projected onto the direction from the void centre to the
   FVS, as a function of cos ($\phi$).
   The angle $\phi$ gives the relative position of the galaxies with
   respect to the FVS direction and is shown in the scheme of the
   upper panel.
   The thick solid line gives the results for SDSS data, with
   uncertainties computed through Jackknife resampling.
   The dotted line corresponds to the simulation.
   The uncertainty region provides a 2--$\sigma$ measure of cosmic variance
   associated with the SDSS volume.
   }
   \label{fig:figure7}
\end{figure}

\section{ANALYSIS IN THE SIMULATION DATA} \label{S_results}

Here we present the main results obtained from the analysis of voids
and FVSs identified in the simulated galaxy catalogue.
In what follows, we develop a measurement for the correlation between
extended regions corresponding to voids and FVSs and analyse their
associated dynamics.

\subsection{Spatial distribution of voids relative to FVSs}

The definition of voids and FVSs results in two types of structures
that are distributed in space occupying somewhat complementary
locations.
In this context, it is worth asking if these structures are related in
a non trivial fashion (i.e., if their distributions are consistent
with a random distribution or if the locations of overdense structures
affect the locations of underdense structures).
In \citet[][]{lares_sparkling_2017}, we show that the spatial
locations of voids are correlated with voids with the same environment
(void--in--void or void--in--cloud types) giving rise to void clumps.
According to this study, clumps of R--type voids show a dynamical
behaviour consistent with divergent flows, produced by a combination
of mutually receding and expanding voids.
On the other hand, clumps of S--type voids exhibit large--scale flows
which are predominantly infalling towards the clump centre.
The global density of these clumps is, in most cases, positive. 
This implies that there must be overdense structures inside clumps
which compensate the void underdensities.
In reference to these ideas, we show in Fig. \ref{fig:figure1} a
visualization of a FVS and its nearby voids for a subset in the
simulation.  We included
A Visualization of a group of void--FVS pairs, including all the voids
(represented as spheres) that have the FVS (in the center of the
Figure) as their closest one.  
Arrows are scaled representations of the void velocity vector, with
dotted segments located inside or behind voids.  
Other FVSs and voids in the neighbourhood are not shown for
simplicity.  
In this context, we argue that large--scale flows play a key role in
the formation of the supercluster--void network, and that
superclusters might be responsible for the global collapse.

Here we focus on the relation between the positions and dynamics of
voids relative to FVSs.
The most common used statistic to quantify the relative distributions
of two types of points is the two--point cross--correlation function,
defined as the excess probability of finding a pair of objects at a
given separation compared to that expected from a random distribution.
However, neither voids nor FVSs can be easily represented by a point
location.
Voids are characterized by a centre and a radius or scale. 
Moreover, FVSs do not have a simple shape and a centre can not be
clearly defined.

We define a procedure to quantify the relative clustering of voids and
FVSs as follows: if the locations of voids and FVSs are both random
and independent, the probability for a randomly placed sphere of to be
in contact with both types of structures should equal the product of
the probability that the sphere touches a void times the probability
of touching a FVS.
These measures certainly depend on the size of the sphere, but if the
distributions of the two types of structures are completely
uncorrelated, the two probabilities are expected to be roughly the
same.
If, on the other hand, they are different, it would indicate a
systematic tendency of either correlation (positive excess), or
anticorrelation (negative excess) with respect to FVSs.

The previously introduced probabilities can be estimated using
classical definitions, where the probability of a randomly placed
sphere of being in contact with a void is approximated by the fraction
of random spheres within the simulation box that contain at least a
part of a void.
The volume fraction occupied by the void is not relevant here, since
this fraction can be studied as a function of the sphere radius.
Then, if $\mathrm{void} \cap \mathrm{FVS}$ denotes the occurrence of a
sphere that is in contact with both a void and a FVS,
\begin{equation} 
   P({\rm void~\cap~FVS}) \simeq \frac{N_{ \mathrm{\rm
   void}~\cap~\mathrm{\rm FVS} }}{N_T}, 
   \label{E_P1}
\end{equation}
\noindent and similarly,
\begin{eqnarray} 
   \label{E_P2} P(\rm void)&\simeq&\frac{N_{\rm void}}{N_T}, \\
   P(\rm FVS)&\simeq&\frac{N_{\rm FVS}}{N_{\rm T}}, 
   \label{E_P3}
\end{eqnarray}
\noindent where $N_V$ and $N_{\rm FVS}$ are the number of spheres that
touch a void and a FVS, respectively.
Also, notice that by definition, $P(\mathrm{void}~\cap~\mathrm{FVS}) =
P({\rm void~|~FVS})~P({\rm FVS})$.
Then, it is equivalent to study $P(\rm void~|~FVS)$ as a function of
$P(\rm void)$.
In the upper panel of Fig.~\ref{fig:figure2} we show these measures
estimated as the fraction of spheres that touch simultaneously an FVS
and a void, and the fraction of spheres that touch a void irrespective
of FVSs.
The upper curve correspond to S--type voids, and the lower curve to
R--type voids, as indicated in the legends.
The circles radii are proportional to the radius of the spheres, as
indicated in the scale at the left.
As it can be seen, there is a clear tendency of S--type voids to be
located closer to FVSs than R--type voids.
Indeed, if a random sphere contains a FVS, the probability of
containing also an S--type void is higher than what should be expected
for a random distribution of voids (represented by the dot-dashed
line).
Conversely, the probability that a FVS and an R--type void are in
contact with the same random sphere is smaller in the full range of
sphere radii, indicating that R--type voids tend to avoid FVSs.
As the sizes of the spheres grow, the difference in probability with
respect to the case of no void clustering decreases since any random
sphere is likely to contain a FVS due to their large volume, and so
the two probabilities become equivalent.
From a scale of $\simeq 40\hmpc$, no significant correlation signal
can be detected.
The probability excess can be writen as:
\begin{equation}
   P({\rm void~|~FVS}) = P({\rm void})\, (1+\xi_{\rm void-FVS}(r)),
   \label{E_P4}
\end{equation}
for a given random sphere radius $r$.

In the bottom panel of this figure we show the void--FVS correlation
function $\xi_{\rm void-FVS}(r)$ for S--type voids (upper curve,
dashed line), R--type voids (bottom curve, solid line), and the
combined sample of voids including both types (dotted line).
It follows the definition of the correlation function given in Eq.
\ref{E_P4}.
It is worth noticing that FVSs represent a small fraction of the total
simulation box volume (nearly 4 per cent, see
\citet{luparello_future_2011}) while voids comprise almost 20 per cent
of the volume.

\subsection{Dynamics of voids relative to FVSs}

In the previous section we showed that there is a tight relation
between the locations of voids of different types and the locations of
FVSs.
Moreover, this geometrical property in the distribution of the largest
scale structures should also have a dynamical counterpart.
The arguments presented in Sec. \ref{S_intro} suggest that the
large--scale flows are closely related to the evolution of the largest
scale structures, and consequently a dynamical connection between
voids and FVSs is expected.
Although the choice of the spherical approximation of voids is
suitable, FVSs can not be described with simple geometrical objects,
which implies that quantities such as relative distance, position or
velocity between those two types of structures are not straightforward
to compute.

Here, we adopt a simple approach that takes into account basic
dynamical principles and allows to study the joint dynamical behaviour
of voids and FVSs.
To analyse the dynamic of FVS--void pairs, as the first step we
identify the nearest FVS for each void.
Since the FVSs have complex shapes, we consider the core region of
each FVS, defined as the volume covered by the highest density cells.
As we mentioned in Sec. \ref{SS_FVS_catalogue}, in the
identification of these structures we have used a physically motivated
threshold on luminosity overdensity \citep{luparello_future_2011},
following the criteria calibrated on simulations by
\citet{dunner_limits_2006}.
According to this procedure, we chose a ratio of luminosity density,
$\rho_{\rm lum}$, relative to the mean satisfying $\rho_{\rm
lum}~/~\overline{\rho}_{\rm lum}>5.5$ to define the boundaries of a
FVS in a 3D grid--averaged smoothed density field estimation.
Thus, since the densest region of the FVSs roughly match the central
region, we define as the central, denser component of a FVS all cells
having $\rho_{\rm lum}~/~\overline{\rho}_{\rm lum}>6.3$ (corresponding
to the median of the luminosity density distribution of FVS cells).
The distance between a void and a FVS is then defined as the distance
between the void centre and the closest cell of the densest core of
the FVS.
For each of the closest void--FVS pair we compute their relative
position and velocity.
To this aim, we define the velocity of the closest FVS cell as the
average of the velocities of the galaxies inside a sphere centred in
the high density cell and with radius equal to the distance between
the void centre and this cell.
The void velocity is then taken relative to their corresponding
closest high density cell.
The angle between these relative position and velocity vectors,
$\theta$, allows to study the flow onto FVSs.

In order to characterize the  void--FVS relative motions, we consider
the preference of relative void velocity vectors towards the direction
to FVSs.
We quantify this coherence by means of a dipole moment estimation,
$\mathcal{D}$, calculated as the average over all void--FVS pairs of
the values of $\cos(\theta)$, weighted by the second order Legendre
polynomial $P_2(x)$ evaluated in $x=\cos(\theta)$:
\begin{equation}
   \mathcal{D} = \frac{1}{N} \sum_{i=1}^{N} \cos(\theta) \, P_2 \big(
   \cos(\theta) \big)  
 \label{E_Dstat}
\end{equation}
It could be argued that S--type voids close to FVSs are likely to show
a stronger infall towards FVS than R-type voids since these are
located more distantly.
To test this hypothesis, we have performed a preliminary analysis by
selecting two subsamples of void--FVS pairs: one comprising only
S--type voids close to a luminous FVS and the other, only by R--type
voids far from low luminosity FVSs. 
The results for these subsamples, shown in the histograms of
\mbox{Fig. \ref{fig:figure3}}, are consistent with a significant infall
pattern of S--types and a lack of coherence of R--types.
In order to visualize how the computed dipole values characterize the
relative infall/outflow patterns, in the inset on top of the figure we
show model distributions of $\cos(\theta)$ with a purely dipole
component.
The scheme in this figure is consistent with distributions of
$\cos(\phi)$ dominated by negative values representing an infall
pattern flowing towards FVS.
We explored the magnitude of the dipole moment as a function of the
relative distance between the void and the FVS, the mass of the
nearest FVS, the mass ratio between the first and second nearest FVSs,
and the distance ratio between the first and second nearest FVSs.

In the upper panel of Fig.~\ref{fig:figure4} we show the dipole moment
$\mathcal{D}$ of the $\cos(\theta)$ distributions in bins of FVS mass
M$_1$ and void--FVS distance d${_1}$.
As it is shown in this figure, the higher negative values of the
dipole (that indicates an infall signal for the pairs) occurs mainly
when the voids are located close to a FVS.
This statistical excess is significant even at distances as large as
$50\hmpc$.
However, it is important to note that besides the low number of
objects, the higher values of $\mathcal{D}$ correspond to the most
massive FVSs.
In grey we show the level curves of the number of pairs FVS--void
which correspond to the 25, 50 and 75 per cent of the sample.
White bins, with a central dot, do not contain any pair.
For the rest of the matrix, we applied a smoothing kernel using a
top--hat window of 2 bins each side (ignoring empty bins).
The color scale indicates the values of the $\mathcal{D}$ statistic
defined in Eq. (\ref{E_Dstat}).
The limits of this scale are the same as the color scale in
Fig.~\ref{fig:figure3}, that gives a picture of the degree of anisotropy
that correspond to the different values of the $\mathcal{D}$
statistic.
In left and right lower panels of Fig.~\ref{fig:figure4} we show the
dipole moment for R--type and S--type voids, respectively.
Box plots correspond to the distributions of $\mathcal{D}$ for each
sample.
There is a stronger infall signal for S--type voids, which is
noticeable up to nearly $25\hmpc$ of void--FVS separation.
Also, in agreement with the results of the previous section, it can be
noticed that S--type voids tend to be located closer to
superstructures than R--type voids.

However, these results are obtained taking into account only the
nearest FVS to each void.
The possible presence of other structures at similar distances lead us
to take into account the influence of at least the second nearest
structure.
To this aim, we compute the relative distance between the void and
their nearest FVS (d$_1$) and the void and their second nearest FVS
(d$_2$), and the ratio of the corresponding FVSs distances and masses.
Following the same scheme of Fig.~\ref{fig:figure4}, in
Fig.~\ref{fig:figure5} we show the dipole moment as a function of the
relative distances and masses.
The data shown in these figures are not exactly the same due to
different mass and distance limits, which produce slightly diferent
box plots.
Here, we can see that if the nearest FVS is remarkably closer to the
void there is a clear infall signal, and this intensifies if the
nearest FVS is much more massive than the second one.  
In this figure we distinguish R-- and S--types showing \mbox{S--type}
voids preferentially located closer to their nearest superstructure.
In Fig.~\ref{fig:figure6} we show the cumulative distributions of
\mbox{void--FVS} distances (bottom and right scales) for
\mbox{R/S--types} and for different void radius intervals
(\mbox{$10.0<R_v/(\rm{Mpc\,h}^{-1})<12.5$} with dotted lines,
\mbox{$12.5<R_v/(Mpc\,h^{-1})<15.0$} with thin solid lines, and
\mbox{$15.0<R_v/(Mpc\,h^{-1})<17.5$} with thick solid lines).
The histograms show the distributions of void radii for the two void
types (upper scale, with arbitrary normalization), with filled bars
for S--types and empty bars for R--types.
The shift in the cumulative distributions are larger considering void
type, indicating that it is not dominated by void size segregation.
Also, in S--type voids we can see that the infall signal grows when
the mass of their nearest FVS (M$_1$) is greater than the mass of the
second one (M$_2$).

\subsection{Galaxies in void shells}

In order to study the influence of the large--scale flows on galaxies
we analyse the velocity field related to the spatial configuration of
the systems. 
To this aim we use SAM central galaxies and identify their nearest FVS
to calculate their relative projected velocity.
Our sample comprises 79403 galaxies, out of which 1996 are located in
R--void shells and 3364 in S--void shells.
We also distinguish between galaxies located in void shells or
elsewhere and we restrict to galaxy--FVS distances in the range
$10-40\hmpc$.
In the upper panel of Fig.~\ref{fig:figure7} we show the mean projected
velocity, V~cos($\theta$), of galaxies in void shells as a function of
their relative angle, $\phi$, with respect to the direction that
connects the void centre and the nearest FVS, as outlined in the
scheme.
The location of a galaxy in the void shell with respect to the FVS
direction is given by $\cos(\phi)$.  Galaxies in the cap facing the
FVS have $\cos(\phi)>0$, and galaxies opposite to the FVS,
$\cos(\phi)<0$, as indicated in the figure axis.
We use a sign convention where negative values of V~cos($\theta$)
correspond to galaxies that are moving towards the FVS.
Negative values of $\cos(\phi)$ correspond to a  galaxy--void--FVS
configuration (as in scheme), while positive values to
void--galaxy--FVS.
It should be noticed that we are using the densest cells in the FVS
definition, so that the cell in the scheme represents the closest cell
in the subset $\rho_{\rm lum}~/~\overline{\rho}_{\rm lum}>6.3$.
It can be appreciated in this figure that galaxies in R--type void
shells may overcome the infall of the void onto the FVS when the
galaxies are located in the direction opposite to the superstructure.
When galaxies are facing directly to the superstructure, they exhibit
the same infall pattern, irrespective of residing in R-- or S--type
void shells.
We notice that this infall is stronger than that of galaxies elsewhere
at the same galaxy--FVS distance due to void dynamics.


\begin{figure} \centering
   \includegraphics[width=\columnwidth]{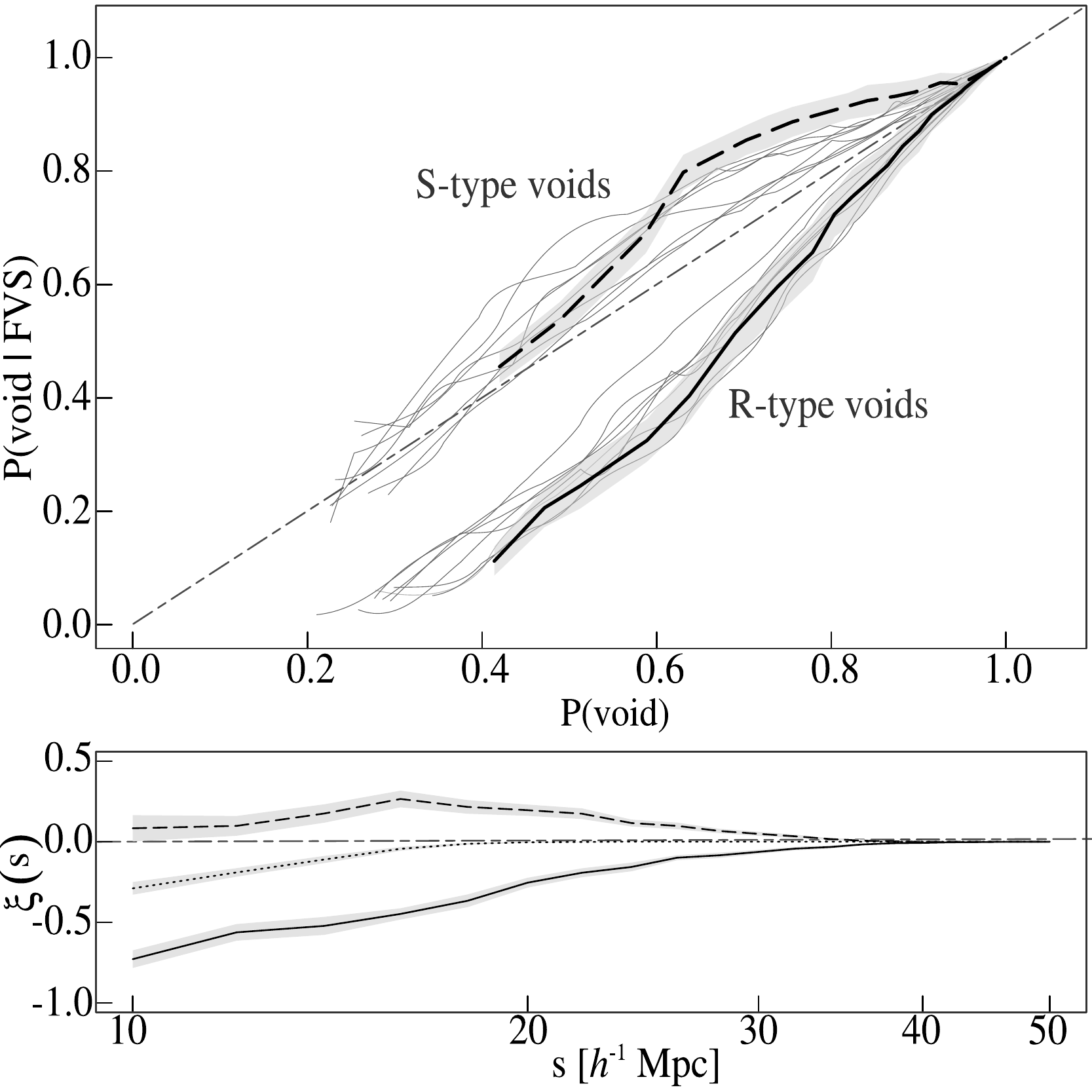} 
   \caption{Spatial distribution of voids with respect to FVSs, in
   the SDSS data.
   Upper panel: Conditional probability of finding a randomly
   placed sphere containing void and FVS volume fractions
   simultaneously, $P({\rm void~|~FVS})$ as a function of the probability that
   the same sphere contains a fraction of void volume $P({\rm void})$.
   The shaded regions indicate Jackknife resampling uncertainties.
   The grey curves correspond to random regions within the simulation
   with the same volume than SDSS sample.
   The bottom panel show the corresponding void--FVS correlations, in
   a similar way to Fig.~\ref{fig:figure2}.}
   \label{fig:figure8}
\end{figure}

\begin{figure*} \centering
   \includegraphics[width=\textwidth]{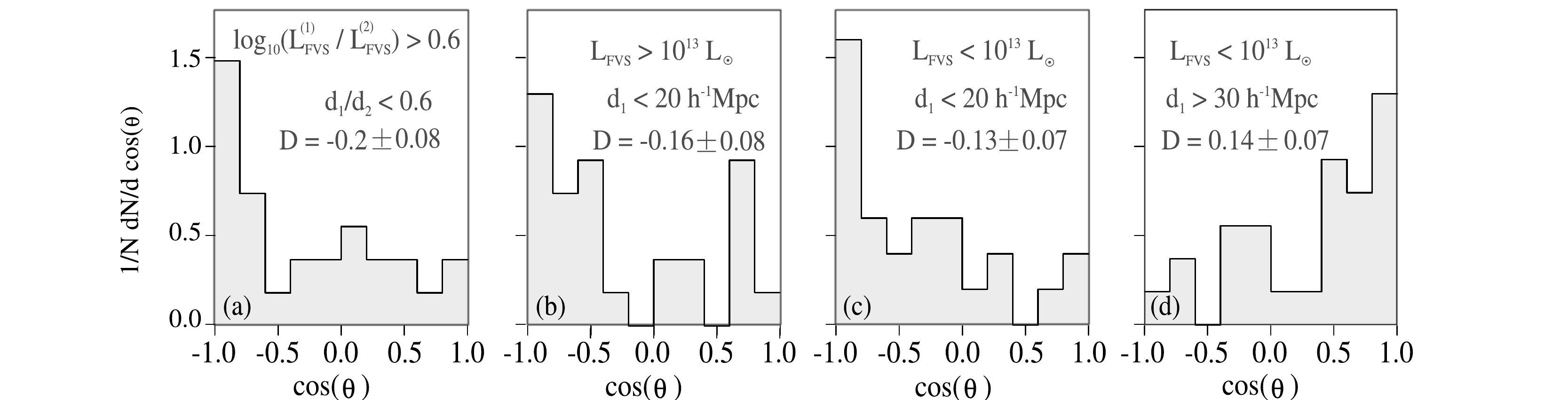} 
   \caption{Histograms of the cos($\theta$) for different
   samples of voids in SDSS.  
   (a) Voids that are at a distance to the
   closest FVS smaller than 60 per cent of the distance
   to the second closest FVS, and also satisfying the ratio of the total
   luminosities of the two
   closest FVS $>0.6$, 
   (b) Voids closer than
   20~$\hmpc$ to FVSs having a total luminosity of at least
   $10^{13}\;L_{\odot}$, 
   (c) Voids closer than 20~$\hmpc$ to FVSs
   having a total luminosity up to $10^{13}\;L_{\odot}$, and
   (d) Voids near FVSs with $L<10^{13}\;L_{\odot}$ separated
   by a distance greater than 30~$\hmpc$.
   We show the values of the dipole moments corresponding to each
   panel, with bootstrap resampling error estimates.
   }
   \label{fig:figure9}
\end{figure*}

\section{OBSERVATIONAL RESULTS} \label{S_observations}

In this section we analyse spatial correlations and dynamics of voids
and FVSs for data in the SDSS catalogue.
As in Fig.~\ref{fig:figure2}, we show in the upper panel of
Fig.~\ref{fig:figure8} the conditional probability of finding a randomly
placed sphere that jointly contain parts of FVSs and voids of S--
(upper curves) or R--types (lower curves), as a function of the
probability that a sphere of the same size contacts at least part of
voids.
We remark the good agreement between observations and the simulation
(Fig.~\ref{fig:figure2}), in spite of the fact that the observational
results correspond to redshift space measurements, due to the large
distances involved.
Since the volume of the SDSS data is much smaller than the volume of
the simulation, we computed the quantities previously defined in Eqs.
\ref{E_P1} to \ref{E_P4}, in regions of the simulation with the same
SDSS volume.
We show with grey curves the corresponding relations for eight
independent regions in the simulation box.
The spread of these curves as a function of $P(void)$ allows a
judgement of the cosmic variance that affects the calculation of the
probabilities.
The grey shaded regions around black curves are the Jackknife
uncertainties for the data.
In the bottom panel of this figure we show the corresponding void--FVS
correlation function for SDSS data.
As it can be seen, the results obtained for SDSS data are consistent
with those of the simulation.
We find that the same general trends obtained in the full simulation
box (Fig.  \ref{fig:figure2}) are suitably reproduced for SDSS data.

In Fig.~\ref{fig:figure9} we show the histograms of $\cos(\theta)$ in
four subsamples to show the effects of mass and distance to the
closest FVSs.
For reference, we show the corresponding values of the dipole moment
$\mathcal{D}$ which should be considered in the context of
Fig.~\ref{fig:figure5}, that gives the same analysis in the simulation
taking advantage of the larger statistics.
In panel (a), we show voids located at a distance to the closest FVS,
smaller than 60 per cent of the distance to the second closest FVS
satisfying a total luminosity ratio of at least $0.6$.
These restrictions take into account the results shown in
Fig.~\ref{fig:figure5} for the simulation box, and corresponds to the
bottom--right corner, where the dipole moment signal has the larger
amplitude.
In this case there is a clear prevalence of negative values of
cos($\theta$), which is consistent with a negative value of the dipole
moment.
In panel (b) we show the histogram of cos($\theta$) for voids closer
than 20$\hmpc$ to FVSs that have a total luminosity of at least
$10^{13}\;L_{\odot}$.
Similarly to the previous subsample, this region is selected from the
upper panel of Fig.~\ref{fig:figure4} corresponding to the bottom--right
corner of the plot.
We also considered two additional restrictions corresponding to voids
closer than 20$\hmpc$ to FVSs and a total luminosity up to
$10^{13}\;L_{\odot}$ (in panel (c) of Fig.~\ref{fig:figure9}), and to
voids associated with luminous FVSs ($L>10^{13}\;L_{\odot}$) separated
by a distance greater than 30$\hmpc$ (panel (d)).
This limiting mass was chosen so that the signal to noise ratio is
high, and taking into account the results in the simulation (Fig.~4)
that indicate that more massive FVSs have a more noticeable dipole
moment. 
We argue that the infall patterns observed in the simulation box are
also reproduced in the SDSS data, despite its much smaller volume and
the linearized velocity field approximation.

In the bottom panel of the Fig.~\ref{fig:figure7} we show the velocity
pattern for the infall of galaxy groups in void shells onto FVSs.
Our sample comprise 189 and 293 groups in R-- and S--type void shells,
respectively, in the $10-40\hmpc$ galaxy--FVS distance range.
The curves show the projected R-S velocity difference of groups in
void shells onto the direction from the void centre to the FVS as a
function of the location of the group.
The angle $\phi$ between the relative position of the group and the
FVS direction is as indicated in the scheme of the
Fig.~\ref{fig:figure7}.
The dotted line is the resulting difference for the simulation.
The solid line corresponds to SDSS data, with error bar indicating
Jackknife resampling uncertainties.
The region around the curve of the simulation gives an estimation of
cosmic variance in order to compare to observations.
It is computed from many randomly placed regions within the simulation
with the same volume than SDSS data, and corresponds to a 2-$\sigma$
uncertainty.


\section{DISCUSSION} \label{S_conclusions} 

In this work we have explored the void--superstructure spatial
cross--correlations and their influence on the large--scale velocity
flows.
To this end, we have studied the velocity field associated with
galaxies in void shells to deepen our understanding of the
supercluster--void network evolution.

Our void and superstructure catalogues are defined according to
\citet{ruiz_clues_2015} and \citet{luparello_future_2011},
respectively. 
Therefore, voids correspond to spherical regions with integrated total
density of 10 percent the mean value up the void radius, and
superstructures are derived from a smoothed luminosity field that
isolates the highest density regions.
These two definitions rely on physical grounds, namely the void sample
has a constant integrated tracer density within void radii, and
superstructures correspond to future virialized systems in the
acelerating $\Lambda$ cold dark matter scenario.

In order to study the spatial correlations between voids and
superstructures, we defined a modified version of the correlation
which is usually applied to point data.
This allows to measure the tendency of voids to be located at a given
distance from a superstructure and detect if they are preferentially
located near FVSs or avoiding them.
We find that while voids--in--clouds (S--types) are preferentially
located near superstructures, voids--in--voids (R--types) are likely
to avoid them.
This is somewhat related to the selection of voids according to their
environment, but we find that it has also dynamical implications when
considered in the context of the large--scale structure, manifested on
the infall of voids which are close to FVS independently of void type.

We also explored galaxies in void shells and how the presence of a
nearby superstructure affects its dynamics.
Galaxies in S-- and R--type void shells within 40~$\hmpc$ infall onto
superstructures except for galaxies opposite to superstructures in
R--type void shells.
Moreover, we find a stronger infall for galaxies residing in void
shells facing the closest superstructure than galaxies at the same
relative distance elsewhere.
We have analysed the similarity of the simulation results and those
inferred from the SDSS galaxy catalogue. In spite
of the SDDS smaller observational volume and linear
velocity approach, and the fact that the Millenium simulation cosmological
parameters differ from more recent observational estimates
\citep{planck_cosmological_2015},
the general trend agreement is encouraging.

Large--scale flows may be analysed into different contexts.
\citep[see for instance][and references therein]{shandarin_multi_2011,
shandarin_cosmic_2012, cautun_evolution_2014}.
In our study, galaxy motions in void shells can be thought as a
combination of shell expansion \citep{paz_clues_2013}, void bulk
motion \citep{lambas_sparkling_2016} and the infall onto
superstructures (this work).
The magnitude of these effects are comparable and need to be taken
into account in simple models for galaxy motions.
Our findings go along with the scenario proposed by
\citet{tully_localvoid_2008} and \citet{tully_laniakea_2014} where the
presence of underdense regions contribute significantly to the
velocity flow of galaxies in the local universe.


\section*{acknowledgements}

This work was partially supported by the Consejo Nacional de
Investigaciones Cient\'{\i}ficas y T\'ecnicas (CONICET), and the
Secretar\'{\i}a de Ciencia y Tecnolog\'{\i}a, Universidad Nacional de
C\'ordoba, Argentina.

Plots were made using \textsc{r} software and postprocessed with Inkscape.
This research has made use of NASA's Astrophysics Data System.

Funding for the SDSS and SDSS-II has been provided by the Alfred P.
Sloan Foundation, the Participating Institutions, the National Science
Foundation, the U.S. Department of Energy, the National Aeronautics
and Space Administration, the Japanese Monbukagakusho, the Max Planck
Society, and the Higher Education Funding Council for England. The
SDSS Web Site is http://www.sdss.org/.
The SDSS is managed by the Astrophysical Research Consortium for the
Participating Institutions. The Participating Institutions are the
American Museum of Natural History, Astrophysical Institute Potsdam,
University of Basel, University of Cambridge, Case Western Reserve
University, University of Chicago, Drexel University, Fermilab, the
Institute for Advanced Study, the Japan Participation Group, Johns
Hopkins University, the Joint Institute for Nuclear Astrophysics, the
Kavli Institute for Particle Astrophysics and Cosmology, the Korean
Scientist Group, the Chinese Academy of Sciences (LAMOST), Los Alamos
National Laboratory, the Max-Planck-Institute for Astronomy (MPIA),
the Max-Planck-Institute for Astrophysics (MPA), New Mexico State
University, Ohio State University, University of Pittsburgh,
University of Portsmouth, Princeton University, the United States
Naval Observatory, and the University of Washington.

\bibliographystyle{mnras}

\newcommand{\noop}[1]{}

\label{lastpage}

\end{document}